\documentclass[a4paper,12pt]{article}
\usepackage{amsfonts}
\usepackage{amsmath}
\newtheorem{conjecture}{Conjecture}
\newtheorem{theorem}{Theorem}
\title{$T$-system and thermodynamic Bethe ansatz equations 
for solvable lattice models associated with superalgebras
\footnote{This is a review paper submitted to the 
proceedings of the workshop:   
`Bilinear Method in the Study of Integrable Systems and Related Topics', 
RIMS, Kyoto, July, 2001 
(URL: http://repository.kulib.kyoto-u.ac.jp/dspace/handle/2433/40854). 
For more details, see the original papers 
\cite{T99-2,T01}.}}  
\author{Zengo Tsuboi\\
Graduate School of Mathematical Sciences, 
University of Tokyo
\footnote{
present address (on January 2010): 
Okayama Institute for Quantum Physics, 
1-9-1 Kyoyama, Okayama 700-0015, Japan}}
\date{}
\begin{document}
\maketitle
\begin{abstract}
 An analytic Bethe ansatz is carried out related to  
 the Lie superalgebra $osp(1|2s)$. 
 We present an eigenvalue formula of a transfer matrix 
 in dressed vacuum form (DVF) labeled by a Young (super) diagram.   
 Remarkable duality among DVFs is found. 
A complete set of transfer matrix functional 
 relations ($T$-system) is proposed as a reduction of a 
 Hirota-Miwa equation. 
 We also derive a thermodynamic Bethe ansatz 
 (TBA) equation from this $T$-system and the 
quantum transfer matrix method. This TBA equation is 
identical to the one from the string hypothesis. 
\end{abstract}
\noindent Journal-ref: RIMS Kokyuroku 1280 (2002) 19-34 \\
URL: 
http://repository.kulib.kyoto-u.ac.jp/dspace/bitstream/2433/42356/1/1280\_03.pdf 
%%%%%%%%%%%%%%%%%%
\section{Introduction}
Solvable lattice models related to 
Lie superalgebras \cite{Ka78} 
have received much attentions
\cite{PS81,KulSk82,Kul86,BS88,DFI90,Sa90,ZBG91,MR94}. 
To construct eigenvalue formulae of transfer matrices 
for such models is an important problem in mathematical 
physics. To achieve this program,  
 the Bethe ansatz has been often used. 
 
Nowadays, there is much literature 
(see for example, \cite{Kul86,Sch87,dVL91,EK92,EKS92,FK93,EK94,Ma95,
Mar95-2,PF96,MR97,ZB97,GM98,JKS98,MNR98,Fr99,Sa99} 
and references therein.) 
on Bethe ansatz analysis 
for solvable lattice models related to Lie superalgebras. 
However, most of it deals only with models related to 
 simple representations like fundamental ones. 
 Only a few people (see for example, \cite{Ma95,PF96})
  tried to deal with 
 more complicated models such as fusion models \cite{KRS81} by 
the Bethe ansatz; while there was no systematic study on this subject. 

To address such situations, we have 
recently executed \cite{T97,T98-1,T98-2,T99-1,T99-2} an 
analytic Bethe ansatz \cite{R83,BR90,KS95,KOS95} systematically 
related to the Lie superalgebras $sl(r+1|s+1),B(r|s),C(s),D(r|s)$ 
 cases. Namely, we have proposed a set of 
 dressed vacuum forms (DVFs) and 
 a class of functional relations ($T$-system) for it. 
Moreover we have also studied  
 thermodynamic Bethe ansatz (TBA) equations \cite{YY69} 
 related to $osp(1|2)$ \cite{ST99,ST00,ST01} and $osp(1|2s)$ \cite{T01} 
 from the point of view of  the string hypothesis \cite{T71,G71} 
and the quantum transfer matrix (QTM) method 
\cite{S85,SI87,K87,SAW90,Kl92,JKS98}. 

In this paper, we briefly review the $T$-system and 
 the TBA equation  related to the Lie superalgebra 
 $osp(1|2s)=B(0|s)$ based on \cite{T99-2,T01}. 
 After a brief review on 
 the Lie superalgebra $osp(1|2s)$, 
 we introduce a QTM for  $osp(1|2s)$ model\cite{Mar95-2} in section 3.
In section 4, 
 we carry out an analytic Bethe ansatz 
based on the Bethe ansatz equation (BAE) 
(\ref{BAE}) and obtain the 
eigenvalue formula  for the QTM.
We  define the dressed vacuum form (DVF) 
 $T_{\lambda \subset \mu}(v)$ labeled by 
a skew-Young (super) diagram $\lambda \subset \mu$ 
as a summation over semi-standard tableaux. 
This DVF has a determinant expression 
(quantum supersymmetric Jacobi-Trudi formula). 
In particular, for a rectangular Young (super) diagram, 
this DVF satisfies 
a kind of Hirota-Miwa equation\cite{H81,M82}. 
By considering a reduction to this equation, 
  we derive the  $osp(1|2s)$ version of the $T$-system. 
  Based on this $T$-system, 
 we derive the TBA equation from the QTM method in section 5. 
 Namely, we consider a dependant variable transformation, 
 and derive the $Y$-system from the $T$-system. 
  Then we transform the $Y$-system with certain analytical 
  conditions into the TBA equation. 
 Moreover we find that 
 this TBA equation coincides with the one from the string hypothesis. 
 This indicates the validity of the string hypothesis for 
 the $osp(1|2s)$ model.  
%%%%%%%%%%%%%%%%%%%%%%%%%%%%%%%%
\section{The Lie superalgebra $osp(1|2s)$}
In this section, we briefly mention the 
Lie superalgebra $B(0|s)=osp(1|2s)$ for $s \in {\mathbb Z}_{\ge 1}$ 
(see for example \cite{Ka78,FJ,MSS}). 

In contrast to other Lie superalgebras, the simple 
root system of $osp(1|2s)$ is unique and  
given as follows (see Figure \ref{dynkin}): 
%%%%%%%%%%%%%%%%%%%%%%%%%%%%%%%%%%%%%%%%%%%%%%%%%%%%%%%%%%%%%%%%%%
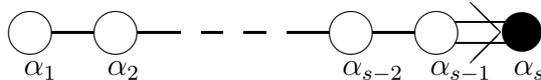
\begin{figure}
    \setlength{\unitlength}{0.8pt}
    \begin{center}
    \begin{picture}(250,50) 
      \put(10,20){\circle{20}}
      \put(20,20){\line(1,0){20}}
      \put(50,20){\circle{20}}
      \put(60,20){\line(1,0){20}}
      \put(90,20){\line(1,0){10}}
      \put(110,20){\line(1,0){10}}
      \put(130,20){\line(1,0){20}}
      \put(160,20){\circle{20}}
      \put(170,20){\line(1,0){20}}
      \put(7,0){$\alpha_{1}$}
      \put(46,0){$\alpha_{2}$}
      \put(156,0){$\alpha_{s-2}$}
      \put(200,20){\circle{20}}
      \put(240,20){\circle*{20}}
      \put(208.8,25){\line(1,0){32.4}}
      \put(208.8,15){\line(1,0){32.4}}
      \put(197,0){$\alpha_{s-1}$}
      \put(236,0){$\alpha_{s}$}
      \put(230,20){\line(-1,1){14.14214}}
      \put(230,20){\line(-1,-1){14.14214}}
  \end{picture}
  \end{center}
  \caption{Dynkin diagram for the Lie superalgebra 
  $B(0|s)=osp(1|2s)$ ($s \ge 1$): white circles denote even roots; 
   a black circle denotes an odd root.}
  \label{dynkin}
\end{figure}
%%%%%%%%%%%%%%%%%%%%%%%%%
\begin{eqnarray}
   && \alpha_{i} = \delta_{i}-\delta_{i+1} 
   \quad {\rm for} \quad i=1,2,\dots,s-1, \nonumber  \\ 
   && \alpha_{s} =\delta_{s}
\end{eqnarray}
where $\delta_{1},\dots,\delta_{s}$ 
are the bases of the dual space of the Cartan subalgebra with the bilinear 
form $(\ |\ )$ such that 
\begin{equation}
 (\delta_{i}|\delta_{j})=-\delta_{i\, j}  
\end{equation}
 $\{\alpha_i \}_{i \ne s}$ are even roots and $\alpha_{s}$ 
is an odd root with $(\alpha_s | \alpha_s)\ne 0$. 
%%%
Let $\lambda \subset \mu$ be a skew-Young (super) diagram labeled by 
the sequences of non-negative integers 
$\lambda =(\lambda_{1},\lambda_{2},\dots)$ and 
$\mu =(\mu_{1},\mu_{2},\dots)$ such that
$\mu_{i} \ge \lambda_{i}: i=1,2,\dots;$  
$\lambda_{1} \ge \lambda_{2} \ge \dots \ge 0$;  
$\mu_{1} \ge \mu_{2} \ge \dots \ge 0$ and 
$\mu^{\prime}=(\mu_{1}^{\prime},\mu_{2}^{\prime},\dots)$ 
be the conjugate of $\mu $. 
In particular, for $\lambda=\phi $, $\mu_{1}\le s$ case, 
the Kac-Dynkin label 
$[b_{1},b_{2},\dots ,  b_{s}]$ is related to 
 the Young (super) diagram with shape 
 $\mu=(\mu_{1},\mu_{2},\dots)$ as follows:  
\begin{eqnarray}
 && b_{i} = \mu_{i}^{\prime}-\mu_{i+1}^{\prime} 
  \qquad {\rm for} \qquad i\in \{1,2,\dots,s-1\}, \nonumber \\ 
 && b_{s} = 2\mu_{s}^{\prime}.  \label{Kac-Dynkin}
\end{eqnarray}  
An irreducible representation 
with the Kac-Dynkin label $[b_{1},b_{2},\dots ,  b_{s}]$
 is finite dimensional if and only if 
\begin{eqnarray}
&& b_{j} \in {\mathbb Z}_{\ge 0} \qquad {\rm for} \qquad 
j \in \{1,2,\dots, s-1\}, \nonumber \\ 
&& b_{s} \in  2{\mathbb Z}_{\ge 0}.
\end{eqnarray}
%%%%%%%%%%%%%%%%%%%%%%%%%%%%%%%%%
\section{$osp(1|2s)$ model and QTM method}
In this section, we introduce an integrable 
spin chain\cite{Mar95-2,MR97} associated with the 
fundamental representation of $osp(1|2s)$, 
and define a QTM. 
The $\check{R}$-matrix\cite{BS88,ZBG91,MR94,MR97} of the model is given as 
\begin{eqnarray}
\check{R}(v)=I+vP-\frac{2v}{2v-g}E,
\end{eqnarray}
where $g=2s+1$; 
$P^{cd}_{ab}=(-1)^{p(a)p(b)}\delta_{ad}\delta_{bc}$; 
$E^{cd}_{ab}=\alpha_{ab}(\alpha^{-1})_{cd}$; 
$a,b,c,d 
\in J=\{1,2,\dots,s,0,\overline{s}, \dots, \overline{2},\overline{1}\}$ 
($ 1 \prec 2 \prec \cdots \prec s \prec 0 \prec 
\overline{s} \prec \cdots \prec \overline{2} \prec \overline{1}$); 
$\alpha$ is $(2s+1)\times (2s+1)$ anti-diagonal matrix whose 
non-zero elements are  
$\alpha_{a,\overline{a}}=1$ for $a \in \{1,2,\dots,s,0\}$ and  
$\alpha_{a,\overline{a}}=-1$ for 
$a \in \{ \overline{s},\overline{s-1},\dots , \overline{1} \} $; 
$\overline{\overline{a}}=a$; 
$p(a)=0$ for $a=0$; $p(a)=1$ for 
$a \in \{1,2,\dots,s\} \sqcup  
\{\overline{s},\dots,\overline{2},\overline{1}\}$. 
The Hamiltonian of the present 
model for the periodic boundary condition is given by 
\begin{eqnarray}
H={\mathcal J}\sum_{k=1}^{L}\left(P_{k,k+1}+\frac{2}{g}E_{k,k+1}\right),
\end{eqnarray}
where 
$L$ is the number of the lattice sites; $P_{k,k+1}$ and 
$E_{k,k+1}$ act nontrivially 
on the $k$ th site and $k+1$ th site. 
There are several formulations of QTM for graded vertex models. 
We consider the case where the transfer matrix is defined as the 
 ordinary trace of a monodromy matrix. 
The QTM is defined as 
\begin{eqnarray}
T^{(1)}_{1}(u,v)={\mathrm Tr}_{j}\prod_{k=1}^{\frac{N}{2}}
 R_{a_{2k},j}(u+iv)\widetilde{R}_{a_{2k-1},j}(u-iv),
 \label{QTM}
\end{eqnarray}
where $R^{cd}_{ab}(v)=\check{R}^{cd}_{ba}(v)$; 
 $\widetilde{R}_{jk}(v)=^{t_{k}}\! \! R_{kj}(v)$ 
($t_{k}$ is the transposition in 
the $k$-th space); $N$ is the Trotter number and assumed to even. 
By using the largest eigenvalue $T^{(1)}_{1}(u_{N},0)$ of 
the QTM (\ref{QTM}), 
the free energy density is expressed as 
\begin{eqnarray}
{\mathcal F}=
-\frac{1}{\beta}\lim_{N\to \infty}\log T^{(1)}_{1}(u_{N},0),
\end{eqnarray}
where $u_{N}=-\frac{{\mathcal J}\beta}{N}$ ($\beta =1/(k_{B}T)$; 
$k_{B}$: the Boltzmann constant; $T$: the temperature). 
From now on, we abbreviate the parameter $u$ in 
$T^{(1)}_{1}(u,v)$. 
%%%%%%%%%%%%%%%%%%%%%%%%%%%%%%%%%%%%%
\section{Analytic Bethe ansatz and $T$-system for QTM}
One can obtain the eigenvalue formulae of the QTM (\ref{QTM}) by 
replacing the vacuum part of the 
DVF for the row-to-row transfer matrix \cite{Mar95-2,MR97}
 with that of the QTM. 
Explicitly we have 
%%%%%%%%
\begin{eqnarray}
T^{(1)}_{1}(v)=\sum_{a\in J}\framebox{$a$}_{v},
\label{DVF}
\end{eqnarray}
where 
the functions $\{\framebox{$a$}_{v}\}_{a\in J}$ are defined as
\begin{eqnarray}
&& \framebox{$a$}_{v}=\psi_{a}(v)
\frac{Q_{a-1}(v+\frac{i}{2}(a+1))Q_{a}(v+\frac{i}{2}(a-2))}
{Q_{a-1}(v+\frac{i}{2}(a-1))Q_{a}(v+\frac{i}{2}a)} 
\nonumber \\
&& \hspace{150pt} \mbox{for} \quad a \in \{1,2,\dots,s\}, 
\nonumber \\
&& \framebox{$0$}_{v}=\psi_{0}(v)
\frac{Q_{s}(v+\frac{i}{2}(s-1))Q_{s}(v+\frac{i}{2}(s+2))}
     {Q_{s}(v+\frac{i}{2}(s+1))Q_{s}(v+\frac{i}{2}s)}, \\  
&& \framebox{$\overline{a}$}_{v}=\psi_{\overline{a}}(v)
\frac{Q_{a-1}(v-\frac{i}{2}(a-2s))Q_{a}(v-\frac{i}{2}(a-2s-3))}
{Q_{a-1}(v-\frac{i}{2}(a-2s-2))Q_{a}(v-\frac{i}{2}(a-2s-1))} 
\nonumber \\
&& \hspace{150pt} \mbox{for} \quad a \in \{1,2,\dots,s\},
\nonumber 
\end{eqnarray}
where $Q_{0}(v):=1$; $\psi_{a}(v)$ is the vacuum part 
\begin{eqnarray}
\psi_{a}(v)=
\left\{
 \begin{array}{lll}
 \zeta_{1}\frac{\phi_{+}(v)\phi_{-}(v+i)\phi_{+}(v-\frac{2s-1}{2}i)}
   {\phi_{+}(v-\frac{2s+1}{2}i)}  &
 \mbox{for} & a=1,\\ 
 \zeta_{a}\phi_{+}(v)\phi_{-}(v)
 &
 \mbox{for} & 2 \preceq a \preceq \overline{2}, \\ 
 \zeta_{\overline{1}}\frac{\phi_{-}(v)\phi_{+}(v-i)\phi_{-}(v+\frac{2s-1}{2}i)}
   {\phi_{-}(v+\frac{2s+1}{2}i)}
 & \mbox{for} & a=\overline{1}, 
 \end{array}
\right.
\label{vac-QTM}
\end{eqnarray}
where $\phi_{\pm}(v)=(v\pm iu)^{\frac{N}{2}}$; 
$\zeta_{a}$ is a phase factor: 
\begin{eqnarray}
\zeta_{a}=
           \left\{
            \begin{array}{lll}
            (-1)^{N-M_{1}} & {\rm if } & a=1 \\ 
             (-1)^{M_{a-1}-M_{a}}
            & {\rm if } & a \in \{2,3,\dots,s\} \\
            1& {\rm if } & a=0 \\
            (-1)^{M_{\overline{a}-1}-M_{\overline{a}}} 
            & {\rm if } & 
            a \in \{\overline{s},\dots,\overline{3},\overline{2}\}\\
            (-1)^{N-M_{1}} & {\rm if } & a=\overline{1},
            \end{array}
           \right.
           \label{pf}
\end{eqnarray}
where $\overline{\overline{a}}=a$.
The complex variables $\{v_{k}^{(a)}\}$ 
are roots of the following  Bethe 
ansatz equation
\begin{eqnarray}
\prod_{j=1}^{N}\left(
\frac{v^{(a)}_{k}-w^{(a)}_{j}+\frac{i}{2}\delta_{a1}}
     {v^{(a)}_{k}-w^{(a)}_{j}-\frac{i}{2}\delta_{a1}}
\right)=
-(-1)^{M_{a-1}-M_{\sigma(a+1)}}
\prod_{d=1}^{s+1}
\frac{Q_{\sigma(d)}(v^{(a)}_{k}+\frac{i}{2}B_{ad})}
     {Q_{\sigma(d)}(v^{(a)}_{k}-\frac{i}{2}B_{ad})},
     \label{BAE}
\end{eqnarray}
where $k\in \{1,2, \dots, M_{a}\}$; $a\in \{1,2, \dots, s\}$; 
$\sigma(d)=d$ for $1\le d \le s$; $\sigma(s+1)=s$; 
$B_{ad}=2\delta_{ad}-\delta_{a,d+1}-\delta_{a,d-1}$;
$Q_{a}(v)=\prod_{k=1}^{M_{a}}(v-v_{k}^{(a)})$; 
$M_{a}\in {\mathbb Z}_{\ge 0}$; $M_{0}=N$. 
The parameter $\sigma$ expresses an effect of 
\symbol{"60}a peculiar two-body self-interaction 
 for the root $\{v_{k}^{(s)}\}$\symbol{"27} \cite{MR97}, 
 which originates from the odd simple root $\alpha_{s}$ with 
 $(\alpha_{s}|\alpha_{s})\ne 0$. 
One may interpret the QTM as a transfer matrix of 
an inhomogeneous vertex model. 
In our case, the inhomogeneity parameters 
$w^{(a)}_{j}\in {\mathbb C}$ 
take the values:
$w^{(a)}_{j}=iu\delta_{a1}$ for $j \in 2{\mathbb Z}_{\ge 1}$; 
$w^{(a)}_{j}=(-iu+\frac{ig}{2})\delta_{a1}$ 
for $j \in 2{\mathbb Z}_{\ge 0}+1$. 
The dress part of the DVF (\ref{DVF}) is free of poles under 
the BAE (\ref{BAE}). 
This is a requirement from the analytic Bethe ansatz \cite{R83}. 

\begin{figure}
    \setlength{\unitlength}{1.5pt}
    \begin{center}
    \begin{picture}(170,350) 
      \put(80,0){\line(1,0){10}}
      \put(90,0){\line(0,1){20}}
      \put(90,20){\line(-1,0){10}}
      \put(80,20){\line(0,-1){20}}
      \put(80,10){\line(1,0){10}}
      \put(83,12){$\overline{1}$}
      \put(83,2){$\overline{1}$}
      \put(74,9){$$}
      %%%%
      \put(80,40){\line(1,0){10}}
      \put(90,40){\line(0,1){20}}
      \put(90,60){\line(-1,0){10}}
      \put(80,60){\line(0,-1){20}}
      \put(80,50){\line(1,0){10}}
      \put(83,52){$\overline{2}$}
      \put(83,42){$\overline{1}$}
      \put(70,47){$$}
      %%%%%%
      \put(40,80){\line(1,0){10}}
      \put(50,80){\line(0,1){20}}
      \put(50,100){\line(-1,0){10}}
      \put(40,100){\line(0,-1){20}}
      \put(40,90){\line(1,0){10}}
      \put(43,92){$\overline{2}$}
      \put(43,82){$\overline{2}$}
      \put(120,80){\line(1,0){10}}
      \put(130,80){\line(0,1){20}}
      \put(130,100){\line(-1,0){10}}
      \put(120,100){\line(0,-1){20}}
      \put(120,90){\line(1,0){10}}
      \put(123,92){$0$}
      \put(123,82){$\overline{1}$}
      \put(110,88){$$}
      %%%%%%
      \put(40,120){\line(1,0){10}}
      \put(50,120){\line(0,1){20}}
      \put(50,140){\line(-1,0){10}}
      \put(40,140){\line(0,-1){20}}
      \put(40,130){\line(1,0){10}}
      \put(43,132){$0$}
      \put(43,122){$\overline{2}$}
      \put(30,128){$$}
      \put(120,120){\line(1,0){10}}
      \put(130,120){\line(0,1){20}}
      \put(130,140){\line(-1,0){10}}
      \put(120,140){\line(0,-1){20}}
      \put(120,130){\line(1,0){10}}
      \put(123,132){$2$}
      \put(123,122){$\overline{1}$}
      \put(109,126){$$}
      %%%%%%
      \put(0,160){\line(1,0){10}}
      \put(10,160){\line(0,1){20}}
      \put(10,180){\line(-1,0){10}}
      \put(0,180){\line(0,-1){20}}
      \put(0,170){\line(1,0){10}}
      \put(3,172){$0$}
      \put(3,162){$0$}
      \put(-10,167){$$}
      \put(80,160){\line(1,0){10}}
      \put(90,160){\line(0,1){20}}
      \put(90,180){\line(-1,0){10}}
      \put(80,180){\line(0,-1){20}}
      \put(80,170){\line(1,0){10}}
      \put(83,172){$2$}
      \put(83,162){$\overline{2}$}
      \put(70,167){$$}
      \put(160,160){\line(1,0){10}}
      \put(170,160){\line(0,1){20}}
      \put(170,180){\line(-1,0){10}}
      \put(160,180){\line(0,-1){20}}
      \put(160,170){\line(1,0){10}}
      \put(163,172){$1$}
      \put(163,162){$\overline{1}$}
      \put(150,167){$$}
      %%%%%%
      \put(40,200){\line(1,0){10}}
      \put(50,200){\line(0,1){20}}
      \put(50,220){\line(-1,0){10}}
      \put(40,220){\line(0,-1){20}}
      \put(40,210){\line(1,0){10}}
      \put(43,212){$2$}
      \put(43,202){$0$}
      \put(30,208){$$}
      \put(120,200){\line(1,0){10}}
      \put(130,200){\line(0,1){20}}
      \put(130,220){\line(-1,0){10}}
      \put(120,220){\line(0,-1){20}}
      \put(120,210){\line(1,0){10}}
      \put(123,212){$1$}
      \put(123,202){$\overline{2}$}
      \put(109,206){$$}
      %%%%%%
      \put(40,240){\line(1,0){10}}
      \put(50,240){\line(0,1){20}}
      \put(50,260){\line(-1,0){10}}
      \put(40,260){\line(0,-1){20}}
      \put(40,250){\line(1,0){10}}
      \put(43,252){$2$}
      \put(43,242){$2$}
      \put(120,240){\line(1,0){10}}
      \put(130,240){\line(0,1){20}}
      \put(130,260){\line(-1,0){10}}
      \put(120,260){\line(0,-1){20}}
      \put(120,250){\line(1,0){10}}
      \put(123,252){$1$}
      \put(123,242){$0$}
      \put(110,248){$$}
      %%%%%%
      \put(80,280){\line(1,0){10}}
      \put(90,280){\line(0,1){20}}
      \put(90,300){\line(-1,0){10}}
      \put(80,300){\line(0,-1){20}}
      \put(80,290){\line(1,0){10}}
      \put(83,292){$1$}
      \put(83,282){$2$}
      \put(74,289){$$}
      %%%%
      \put(80,320){\line(1,0){10}}
      \put(90,320){\line(0,1){20}}
      \put(90,340){\line(-1,0){10}}
      \put(80,340){\line(0,-1){20}}
      \put(80,330){\line(1,0){10}}
      \put(83,332){$1$}
      \put(83,322){$1$}
      \put(74,329){$$}
      %%%%%%%%%%%%
      \put(52,78){\vector(3,-2){26}}
      \put(62,73){\scriptsize{$(1,5)$}}
      %%%
      \put(12,158){\vector(3,-2){26}}
      \put(22,153){\scriptsize{$(2,4)$}}
      \put(92,158){\vector(3,-2){26}}
      \put(102,153){\scriptsize{$(1,5)$}}
      %%%
      \put(52,198){\vector(3,-2){26}}
      \put(62,193){\scriptsize{$(2,4)$}}
      \put(132,198){\vector(3,-2){26}}
      \put(142,193){\scriptsize{$(1,5)$}}
      %%%
      \put(92,278){\vector(3,-2){26}}
      \put(102,273){\scriptsize{$(2,3)$}}
      %%%%%
      \put(118,78){\vector(-3,-2){26}}
      \put(92,73){\scriptsize{$(2,2)$}}
      %%%
      \put(78,158){\vector(-3,-2){26}}
      \put(52,153){\scriptsize{$(2,1)$}}
      \put(158,158){\vector(-3,-2){26}}
      \put(132,153){\scriptsize{$(1,0)$}}
      %%%
      \put(38,198){\vector(-3,-2){26}}
      \put(12,193){\scriptsize{$(2,1)$}}
      \put(118,198){\vector(-3,-2){26}}
      \put(92,193){\scriptsize{$(1,0)$}}
      %%%
      \put(78,278){\vector(-3,-2){26}}
      \put(52,273){\scriptsize{$(1,0)$}}
      %%%%%%%
      %
      \put(85,37){\vector(0,-1){14}}
      \put(87,29){\scriptsize{$(1,3)$}}
      %%%
      \put(45,117){\vector(0,-1){14}}
      \put(47,109){\scriptsize{$(2,2)$}}
      \put(125,117){\vector(0,-1){14}}
      \put(127,109){\scriptsize{$(2,1)$}}
      %%%
      \put(45,237){\vector(0,-1){14}}
      \put(47,229){\scriptsize{$(2,3)$}}
      \put(125,237){\vector(0,-1){14}}
      \put(127,229){\scriptsize{$(2,4)$}}
      %%%
      \put(85,317){\vector(0,-1){14}}
      \put(87,309){\scriptsize{$(1,2)$}}
      %%%%%%%
      \put(54,120){\vector(3,-1){62}}
      \put(84,112){\scriptsize{$(1,5)$}}
      \put(116,240){\vector(-3,-1){62}}
      \put(77,237){\scriptsize{$(1,0)$}}
  \end{picture}
  \end{center}
  \caption{The Bethe-strap structure of  
  $T^{(1)}_{2}(v)$  for 
   $osp(1|4)$:  
 The topmost tableau corresponds to the 
 \symbol{96}highest weight \symbol{39}, 
 which is called the {\em top term}.
 This term carries the $osp(1|4)$ weight $2\delta_{1}$}
  \label{best2}
\end{figure}
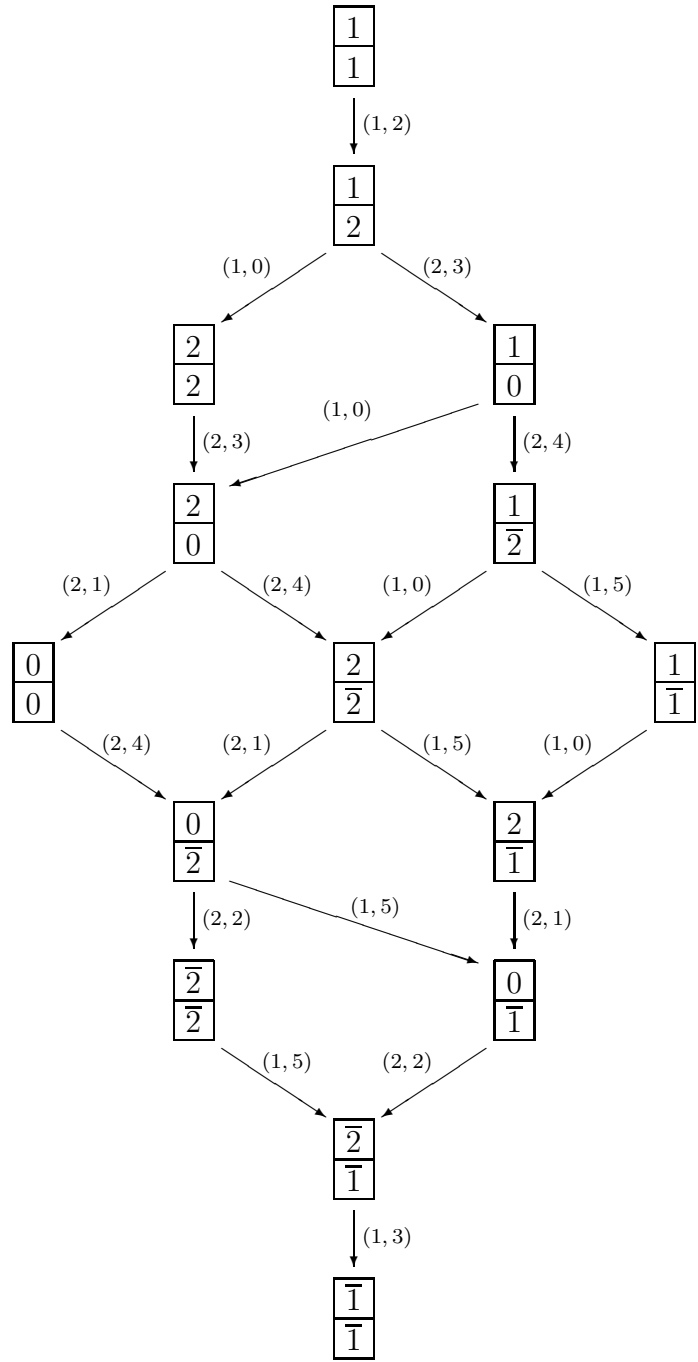

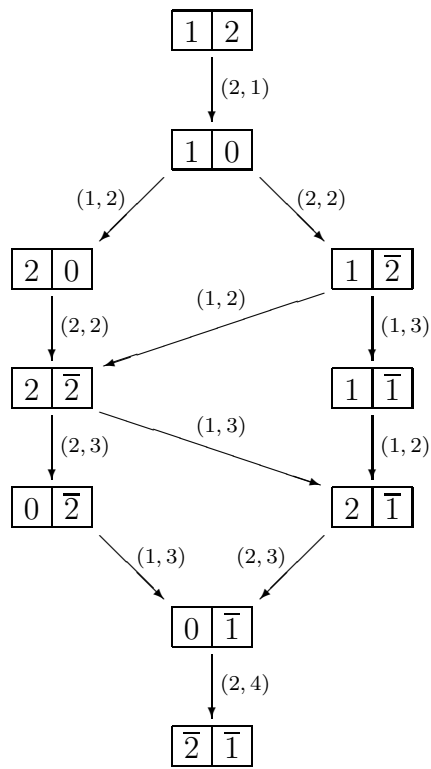
\begin{figure}
    \setlength{\unitlength}{1.5pt}
    \begin{center}
    \begin{picture}(100,200) 
      \put(40,0){\line(1,0){20}}
      \put(60,0){\line(0,1){10}}
      \put(60,10){\line(-1,0){20}}
      \put(40,10){\line(0,-1){10}}
      \put(50,0){\line(0,1){10}}
      \put(43,2){$\overline{2}$}
      \put(53,2){$\overline{1}$}
      \put(32,3){$$}
      %%%%%%% 
      \put(40,30){\line(1,0){20}}
      \put(60,30){\line(0,1){10}}
      \put(60,40){\line(-1,0){20}}
      \put(40,40){\line(0,-1){10}}
      \put(50,30){\line(0,1){10}}
      \put(43,32){$0$}
      \put(53,32){$\overline{1}$}
      \put(32,33){$$}
      %%%
      \put(0,60){\line(1,0){20}}
      \put(20,60){\line(0,1){10}}
      \put(20,70){\line(-1,0){20}}
      \put(0,70){\line(0,-1){10}}
      \put(10,60){\line(0,1){10}}
      \put(3,62){$0$}
      \put(13,62){$\overline{2}$}
      \put(-7,63){$$}
      \put(80,60){\line(1,0){20}}
      \put(100,60){\line(0,1){10}}
      \put(100,70){\line(-1,0){20}}
      \put(80,70){\line(0,-1){10}}
      \put(90,60){\line(0,1){10}}
      \put(83,62){$2$}
      \put(93,62){$\overline{1}$}
      %%%
      %%%
      \put(0,90){\line(1,0){20}}
      \put(20,90){\line(0,1){10}}
      \put(20,100){\line(-1,0){20}}
      \put(0,100){\line(0,-1){10}}
      \put(10,90){\line(0,1){10}}
      \put(3,92){$2$}
      \put(13,92){$\overline{2}$}
      \put(-7,93){$$}
      \put(80,90){\line(1,0){20}}
      \put(100,90){\line(0,1){10}}
      \put(100,100){\line(-1,0){20}}
      \put(80,100){\line(0,-1){10}}
      \put(90,90){\line(0,1){10}}
      \put(83,92){$1$}
      \put(93,92){$\overline{1}$}
      %%%
      %%%
      \put(0,120){\line(1,0){20}}
      \put(20,120){\line(0,1){10}}
      \put(20,130){\line(-1,0){20}}
      \put(0,130){\line(0,-1){10}}
      \put(10,120){\line(0,1){10}}
      \put(3,122){$2$}
      \put(13,122){$0$}
      \put(-7,123){$$}
      \put(80,120){\line(1,0){20}}
      \put(100,120){\line(0,1){10}}
      \put(100,130){\line(-1,0){20}}
      \put(80,130){\line(0,-1){10}}
      \put(90,120){\line(0,1){10}}
      \put(83,122){$1$}
      \put(93,122){$\overline{2}$}
      %%%%%%%%
      \put(40,150){\line(1,0){20}}
      \put(60,150){\line(0,1){10}}
      \put(60,160){\line(-1,0){20}}
      \put(40,160){\line(0,-1){10}}
      \put(50,150){\line(0,1){10}}
      \put(43,152){$1$}
      \put(53,152){$0$}
      \put(32,153){$$}
      %%%
      \put(40,180){\line(1,0){20}}
      \put(60,180){\line(0,1){10}}
      \put(60,190){\line(-1,0){20}}
      \put(40,190){\line(0,-1){10}}
      \put(50,180){\line(0,1){10}}
      \put(43,182){$1$}
      \put(53,182){$2$}
      \put(32,183){$$}
      %%%%%%%%%%%%
      \put(10,88){\vector(0,-1){16}}
      \put(12,79){\scriptsize{$(2,3)$}}
      \put(90,88){\vector(0,-1){16}}
      \put(92,79){\scriptsize{$(1,2)$}}
      \put(10,118){\vector(0,-1){16}}
      \put(12,109){\scriptsize{$(2,2)$}}
      \put(90,118){\vector(0,-1){16}}
      \put(92,109){\scriptsize{$(1,3)$}}
      %%%%
      \put(50,28){\vector(0,-1){16}}
      \put(52,19){\scriptsize{$(2,4)$}}
      \put(50,178){\vector(0,-1){16}}
      \put(52,169){\scriptsize{$(2,1)$}}
      %%%%%%%%%%%
      \put(22,58){\vector(1,-1){16}}
      \put(31,51){\scriptsize{$(1,3)$}}
      %%%%
      \put(78,58){\vector(-1,-1){16}}
      \put(56,51){\scriptsize{$(2,3)$}}
      %%%%%%%%%
      \put(62,148){\vector(1,-1){16}}
      \put(71,141){\scriptsize{$(2,2)$}}
      %%%%
      \put(38,148){\vector(-1,-1){16}}
      \put(16,141){\scriptsize{$(1,2)$}}
      %%%%%%%%%%%
      \put(22,89){\vector(3,-1){55}}
      \put(46,84){\scriptsize{$(1,3)$}}
      %%%%%%%%%
      \put(78,119){\vector(-3,-1){55}}
      \put(46,116){\scriptsize{$(1,2)$}}
      %%%%%%%%%%
  \end{picture}
  \end{center}
  \caption{The Bethe-strap structure of 
  $T^{(2)}_{1}(v)$ for $osp(1|4)$:  
 The topmost  tableau corresponds to the 
 \symbol{96}highest weight \symbol{39}, 
 which is called the {\em top term}.
  This term carries the $osp(1|4)$ weight $\delta_{1}+\delta_{2}$}
  \label{best3}
\end{figure}

%%%%%%%%%%%%%%%%%%
Now we will present a DVF $T_{\lambda \subset \mu}(v)$ 
for a \symbol{"60}fusion QTM\symbol{"27}. 
 We can derive the explicit expression of  $T_{\lambda \subset \mu}(v)$
by modifying the vacuum part of the DVF in Ref. \cite{T99-2} 
 so that the vacuum part is compatible with the left 
 hand side of the BAE (\ref{BAE}). 
We assign coordinates $(i,j)\in {\mathbb Z}^{2}$ 
on the skew-Young  (super) diagram $\lambda \subset \mu$ 
such that the row index $i$ increases as we go downwards and the column 
index $j$ increases as we go from the left to the right and that 
$(1,1)$ is on the top left corner of $\mu$.
We define an admissible tableau $b$ 
on the skew-Young (super) diagram 
$\lambda \subset \mu$ as a set of elements $b(i,j)\in J$ 
 labeled by the coordinates 
$(i,j)$ mentioned above, with the following rule 
(admissibility conditions).
\begin{eqnarray}
 b(i,j) \prec b(i,j+1),  \qquad  
 b(i,j) \preceq b(i+1,j). \label{adm}
\end{eqnarray}
Let $B(\lambda \subset \mu)$ be 
the set of admissible tableaux 
 on $\lambda \subset \mu$. 
For any skew-Young (super) diagram $\lambda \subset \mu$, 
define $T_{\lambda \subset \mu}(v)$ as follows
\begin{equation}
 T_{\lambda \subset \mu}(v)=
\sum_{b \in B(\lambda \subset \mu)}
\prod_{(j,k) \in (\lambda \subset \mu)}
\framebox{$b(j,k)$}_{v-\frac{i}{2}(-\mu_{1}+\mu_{1}^{\prime}-2j+2k)}	
\label{DVF-tb},
\end{equation}
where the product is taken over the coordinates $(j,k)$ on
 $\lambda \subset \mu$. 
 Let $T_{m}^{(a)}(v):=T_{(a^{m})}(v)$. 
 The following determinant formula 
 (quantum supersymmetric Jacobi-Trudi formula) 
 should be
 valid (cf. \cite{KOS95}).
\begin{eqnarray}
T_{\lambda \subset \mu}(v)&=&{\rm det}_{1 \le j,k \le \mu_{1}^{\prime}}
    (T^{(\mu_{k}-\lambda_{j}+j-k)}_{1}
    (v- 
    \nonumber \\ 
  && \qquad \frac{i}{2}
    (-\mu_{1}+\mu_{1}^{\prime}+
    \mu_{k}^{\prime}+\lambda_{j}^{\prime}-j-k+1)))
	\label{Jacobi-Trudi} .
\end{eqnarray}
%%%%%%%%%%%%%%%%%%%%%%%%%%%%%%%%%%%%%
  We may think of (\ref{DVF-tb}) as an $osp(1|2s)$ version of the 
 Bazhanov and Reshetikhin's eigenvalue formula \cite{BR90}. 
%
%%%%%%%%%%%%%
In particular, for $\lambda =\phi$, $\mu_{1}\le s$ case, 
 the \symbol{"60}top term\symbol{"27} of $T_{\mu}(v)$ will be 
 the term corresponding to the tableau 
 $b(i,j)=j$ ($1 \le i \le \mu_{j}^{\prime}$, $1 \le j \le s$).  
 This term carries the $osp(1|2s)$ weight with the 
 Kac-Dynkin label (\ref{Kac-Dynkin}) (in the sense in Ref. \cite{KS95}). 
 %%%%%%%%%%
 DVFs have so called 
{\em Bethe-strap} structures 
\cite{KS95} and we confirmed, for several examples,
  that $T_{\lambda \subset \mu}(v)$ coincides with 
 the Bethe-strap of the minimal connected component which 
 includes the top term  
 as the examples in 
 Figure \ref{best1}, Figure \ref{best2} and Figure \ref{best3}. 
 $T_{\lambda \subset \mu}(v)$ may be viewed as a prototype of 
a \symbol{"60}$q$-supercharacter\symbol{"27} (cf. \cite{FM01}).  
%%%%%%%%%%%%%%%%%%%%%%%%%%%%%%%%%%%%%%%%
\begin{figure}
    \setlength{\unitlength}{1.5pt}
    \begin{center}
    \begin{picture}(180,40) 
      \put(-8,3){$$}
      \put(0,0){\line(1,0){10}}
      \put(10,0){\line(0,1){10}}
      \put(10,10){\line(-1,0){10}}
      \put(0,10){\line(0,-1){10}}
      \put(3,3){\scriptsize{$1$}}
      \put(12,5){\vector(1,0){20}}
      \put(15,9){\scriptsize{$(1,1)$}}
      \put(33,3){$$}
      \put(40,0){\line(1,0){10}}
      \put(50,0){\line(0,1){10}}
      \put(50,10){\line(-1,0){10}}
      \put(40,10){\line(0,-1){10}}
      \put(43,3){\scriptsize{$2$}}
      \put(52,5){\vector(1,0){20}}
      \put(55,9){\scriptsize{$(2,2)$}}
      \put(73,3){$$}
      \put(80,0){\line(1,0){10}}
      \put(90,0){\line(0,1){10}}
      \put(90,10){\line(-1,0){10}}
      \put(80,10){\line(0,-1){10}}
      \put(83,3){\scriptsize{$0$}}
      \put(92,5){\vector(1,0){20}}
      \put(95,9){\scriptsize{$(2,3)$}}
      \put(113,3){$$}
      \put(120,0){\line(1,0){10}}
      \put(130,0){\line(0,1){10}}
      \put(130,10){\line(-1,0){10}}
      \put(120,10){\line(0,-1){10}}
      \put(123,3){\scriptsize{$\overline{2}$}}
      \put(132,5){\vector(1,0){20}}
      \put(135,9){\scriptsize{$(1,4)$}}
      \put(153,3){$$}
      \put(160,0){\line(1,0){10}}
      \put(170,0){\line(0,1){10}}
      \put(170,10){\line(-1,0){10}}
      \put(160,10){\line(0,-1){10}}
      \put(163,3){\scriptsize{$\overline{1}$}}
  \end{picture}
  \end{center}
  \caption{The  Bethe-strap structure of  
   $T^{(1)}_{1}(v)$  for 
   $osp(1|4)$:  
 The pair $(a,b)$ denotes the common pole 
 $v_{k}^{(a)}-\frac{i}{2}b$ of the pair   
 of the tableaux connected by the arrow.   
 This common pole vanishes under the BAE (\ref{BAE}).
 The leftmost tableau corresponds to the 
 \symbol{96}highest weight \symbol{39}, 
 which is called the {\em top term}. 
 This term carries the $osp(1|4)$ weight $\delta_{1}$.} 
  \label{best1}
\end{figure}
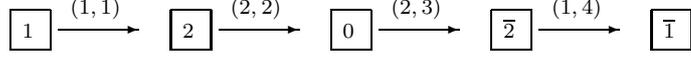
%%%%%%%%%%%%%%%%%%%%%
%%%%%%%%%%%%%%%%%%%%%%%%%%%%%%%%%%%%%%%%%%
%%%%%%%%%%%%%%%%%%%%%%%%%
%
%%%%%%%%%%%%%%%%%%%%%%%%%%%%%%%

Now we introduce the functional relations among DVFs. 
The following relation follows from the determinant formula 
(\ref{Jacobi-Trudi}). 
\begin{eqnarray}
&& \hspace{-20pt}
T^{(a)}_{m}(v+\frac{i}{2})T^{(a)}_{m}(v-\frac{i}{2})
=T^{(a)}_{m+1}(v)T^{(a)}_{m-1}(v)+T^{(a-1)}_{m}(v)T^{(a+1)}_{m}(v)
\label{Hirota-Miwa}, 
\end{eqnarray}
where $a,m \in {\mathbb Z}_{\ge 1}$. 
This functional relation is a kind of 
Hirota-Miwa equation \cite {H81,M82} 
and can be 
proved by the Jacobi identity. 
The following theorem follows from the admissible condition (\ref{adm}).  
\begin{theorem}\label{cons}
$T_{\lambda \subset \mu}(v) =0$ if 
 $\lambda \subset \mu$ contains $m \times a$ rectangular subdiagram 
 ($m$: the number of row, $a$: the number of column) with 
$a \in {\mathbb Z}_{\ge 2s+2}$ and $m \in {\mathbb Z}_{\ge 1}$.
In particular, we have 
\begin{eqnarray}
T_{m}^{(a)}(v)=0 \qquad {\rm if } \quad 
 a \in {\mathbb Z}_{\ge 2s+2} \quad {\rm and}
  \quad m \in {\mathbb Z}_{\ge 1}.
\end{eqnarray}
\end{theorem}
There is a remarkable duality for $T_{m}^{(a)}(v)$. 
\begin{theorem}\label{dual}
For any $a \in \{1,\dots,s \}$ and 
$m \in {\mathbb Z}_{\ge 0}$, we have 
\begin{eqnarray}
 T^{(a)}_{m}(v)={\mathcal M}_{m}^{(a)}(v)T^{(2s-a+1)}_{m}(v),
\end{eqnarray}
where ${\mathcal M}_{m}^{(a)}(v)$ is given as 
\begin{eqnarray}
{\mathcal M}_{m}^{(a)}(v)&=&
\prod_{j=1}^{m}
\left\{
\frac{\psi_{1}(v-\frac{i}{2}(m-a-2j+2))}
{\psi_{1}(v-\frac{i}{2}(m-2s+a-2j+1))}
\right. \nonumber \\ 
&& \left. \times 
 \frac{\prod_{k=2}^{a}\psi_{2}(v-\frac{i}{2}(m-a-2j+2k))}
 {\prod_{k=2}^{2s-a+1}\psi_{2}(v-\frac{i}{2}(m-2s+a-2j+2k-1))}
\right\}.
\end{eqnarray}
\end{theorem}
%%%%%%%%%%%%%%%%%
For $a \in \{1,2,\dots,s\}$ and $m \in {\mathbb Z}_{\ge 1}$, we 
 define a normalization function
\begin{eqnarray}
{\mathcal N}^{(a)}_{m}(v)=
 \frac{\prod_{j=1}^{m}\prod_{k=1}^{a}
  \phi_{-}(v-\frac{m-a-2j+2k}{2}i)\phi_{+}(v-\frac{m-a-2j+2k}{2}i)}{
  \phi_{-}(v-\frac{m-a}{2}i)\phi_{+}(v+\frac{m-a}{2}i)}.
\end{eqnarray}
 We reset $T^{(a)}_{m}(v)/{\mathcal N}^{(a)}_{m}(v)$ to 
 $T^{(a)}_{m}(v)$, where $T^{(a)}_{m}(v)$ is defined by 
 (\ref{DVF-tb}). 
By using the Theorem \ref{cons},\ref{dual}, 
we can obtain the $T$-system as a
 reduction of the Hirota-Miwa equation (\ref{Hirota-Miwa}).
\begin{eqnarray}
&& \hspace{-20pt}
T^{(a)}_{m}(v+\frac{i}{2})T^{(a)}_{m}(v-\frac{i}{2})
=T^{(a)}_{m+1}(v)T^{(a)}_{m-1}(v)+T^{(a-1)}_{m}(v)T^{(a+1)}_{m}(v)
\nonumber \\ 
&& \hspace{130pt} {\rm for} \quad a \in {1,2,\dots,s-1}, 
\label{T-system} \\ 
&& \hspace{-20pt}
T^{(s)}_{m}(v+\frac{i}{2})T^{(s)}_{m}(v-\frac{i}{2})
=T^{(s)}_{m+1}(v)T^{(s)}_{m-1}(v)+
g^{(s)}_{m}(v)T^{(s-1)}_{m}(v)T^{(s)}_{m}(v), \nonumber 
\end{eqnarray}
where 
\begin{eqnarray}
T^{(a)}_{0}(v)&=&\phi_{-}(v+\frac{a}{2}i)\phi_{+}(v-\frac{a}{2}i) 
\quad {\rm for} \quad a \in {\mathbb Z}_{\ge 1},\nonumber \\
T^{(0)}_{m}(v)&=&\phi_{-}(v-\frac{m}{2}i)\phi_{+}(v+\frac{m}{2}i)
\quad {\rm for} \quad m \in {\mathbb Z}_{\ge 1}, \\
g^{(s)}_{m}(v)&=&
 \frac{\phi_{-}(v+\frac{m+s+1}{2}i)\phi_{+}(v-\frac{m+s+1}{2}i)}
 {\phi_{-}(v+\frac{m+s}{2}i)\phi_{+}(v-\frac{m+s}{2}i)}
\quad {\rm for} \quad m \in {\mathbb Z}_{\ge 1}.\nonumber 
\end{eqnarray}
For $s=1$, $g^{(1)}_{m}(v)T^{(0)}_{m}(v)$ coincides with 
the function $T^{(0)}_{m}(v)$ in Ref. \cite{ST00}.
Since the dress part of the DVF $T^{(a)}_{m}(v)$ is same as  
the row-to-row case, this 
functional equation (\ref{T-system})
 has essentially the same form as the $osp(1|2s)$ 
$T$-system in Ref.\cite{T99-2}. 
%%%%%%%%%%%%%%%%%%%%%
\section{TBA equation}
For $m \in {\mathbb Z}_{\ge 1}$, we define the $Y$-functions:
\begin{eqnarray}
 Y^{(a)}_{m}(v)&=&
 \frac{T^{(a)}_{m+1}(v)T^{(a)}_{m-1}(v)}{T^{(a-1)}_{m}(v)T^{(a+1)}_{m}(v)}
 \quad {\rm for} 
 \quad a \in \{1,2,\dots,s-1\}, \nonumber \\
 Y^{(s)}_{m}(v)&=&
 \frac{T^{(s)}_{m+1}(v)T^{(s)}_{m-1}(v)}
 {g^{(s)}_{m}(v)T^{(s-1)}_{m}(v)T^{(s)}_{m}(v)}.
 \label{Y-fun}
\end{eqnarray}
By using the $T$-system (\ref{T-system}), one can show that 
the $Y$-functions satisfy the following $Y$-system:
\begin{eqnarray}
 Y^{(a)}_{m}(v+\frac{i}{2})Y^{(a)}_{m}(v-\frac{i}{2})&=&
 \frac{(1+Y^{(a)}_{m+1}(v))(1+Y^{(a)}_{m-1}(v))}
 {\prod_{d=1}^{s}(1+(Y_{m}^{(d)}(v))^{-1})^{I_{ad}}},
 \label{Y-sys}
 \end{eqnarray}
where $Y^{(a)}_{0}(v)=0$, $a \in \{1,2,\dots,s\}$ and 
$m \in {\mathbb Z}_{\ge 1}$;
$I_{ad}=\delta_{a,d-1}+\delta_{a,d+1}+\delta_{a d}\delta_{a s}$. 
A numerical analysis for finite $N,u,s$ indicates that 
a two-string solution (for every color) in the sector 
$N=M_{1}=M_{2}=\cdots =M_{s}$ of the BAE (\ref{BAE}) provides 
the largest eigenvalue of the QTM (\ref{QTM}) at $v=0$.
Moreover, we expect the following conjecture is valid for 
this two-string solution. 
\begin{conjecture}\label{conj}
For small $u$ ($|u|\ll 1$) and $a \in \{1,2,\dots,s\}$,
every zero of $T^{(a)}_{m}(v)$ is located outside of 
the physical strip 
${\rm Im}v \in [-\frac{1}{2},\frac{1}{2}]$.
\end{conjecture}
Based on this conjecture, 
we shall establish the ANZC property in some domain 
for the $Y$-functions (\ref{Y-fun}) 
to transform the $Y$-system (\ref{Y-sys}) 
 to nonlinear integral equations. 
Here ANZC means Analytic NonZero and Constant asymptotics in the 
limit $|v| \to \infty$. 
One can show that the $Y$-function has the following asymptotic value 
\begin{eqnarray}
\lim_{|v| \to \infty}Y^{(a)}_{m}(v)=\frac{m(g+m)}{a(g-a)},
\label{limit}
\end{eqnarray}
which is identified to the solution 
of the constant $Y$-system
\begin{eqnarray}
(Y_{m}^{(a)})^{2}=
\frac{(1+Y_{m-1}^{(a)})(1+Y_{m+1}^{(a)})}
 {\prod_{d=1}^{s}(1+(Y_{m}^{(d)})^{-1})^{I_{ad}}}, 
 \label{const-Y}
\end{eqnarray}
where $Y_{0}^{(a)}:=0$,
 $a \in \{1,2,\dots, s\}$ and $m \in {\mathbb Z}_{\ge 1}$. 
From the Conjecture \ref{conj} and (\ref{limit}), we find that the 
functions $1+Y_{m}^{(a)}(v)$, $1+(Y_{m}^{(a)}(v))^{-1}$ 
 in the domain ${\rm Im}v \in [-\delta,\delta]$ 
($0<\delta \ll 1$) and  
$Y^{(a)}_{m}(v)$ for $(a,m)\ne (1,1)$ in the domain  
${\rm Im}v \in [-\frac{1}{2},\frac{1}{2}]$ (physical strip) 
have the ANZC property. On the other hand, 
$Y^{(1)}_{1}(v)$ has zeros of order $N/2$ 
at $\pm i(\frac{1}{2}-u)$ if $u>0$ (${\mathcal J}<0$), 
poles of order $N/2$ 
at $\pm i(\frac{1}{2}+u)$ if $u<0$ (${\mathcal J}>0$) in the physical strip. 
Then we must modify $Y^{(1)}_{1}(v)$ as 
\begin{eqnarray}
\hspace{-10pt}
\widetilde{Y}^{(a)}_{m}(v)=Y^{(a)}_{m}(v)\left\{
\tanh\frac{\pi}{2}(v+i(\frac{1}{2} \pm u))\tanh\frac{\pi}{2}
(v-i(\frac{1}{2} \pm u))\right\}^{\pm \frac{N\delta_{a1}\delta_{m1}}{2}}
\hspace{-20pt} ,
\end{eqnarray}
where the sign $\pm $ is identical to that of $-u$. 
Taking note on the relation
\begin{eqnarray}
\tanh\frac{\pi}{4}(v+i)\tanh\frac{\pi}{4}(v-i)=1,
\end{eqnarray}
one can modify the lhs of the $Y$-system (\ref{Y-sys}) as
\begin{eqnarray} 
&& \widetilde{Y}^{(a)}_{m}(v-\frac{i}{2})\widetilde{Y}^{(a)}_{m}(v+\frac{i}{2})
 =\frac{(1+Y^{(a)}_{m+1}(v))(1+Y^{(a)}_{m-1}(v))}
 {\prod_{d=1}^{s}(1+(Y_{m}^{(d)}(v))^{-1})^{I_{ad}}}, 
 \label{modi-Y-sys} \\
&& \hspace{120pt}{\rm for} \quad m \in {\mathbb Z}_{\ge 1} 
\quad {\rm and} \quad a\in \{1,2,\dots,s\}.\nonumber 
\end{eqnarray}
Now that the ANZC property has been established for the $Y$-system, 
we can transform (\ref{modi-Y-sys}) into a system of 
nonlinear integral equations 
by a standard procedure. 
\begin{eqnarray}
\log Y_{m}^{(a)}(v)&=&
\mp \frac{N\delta_{a1}\delta_{m1}}{2}
\log \left\{
\tanh\frac{\pi}{2}(v+i(\frac{1}{2} \pm u))\tanh\frac{\pi}{2}
(v-i(\frac{1}{2} \pm u))
\right\} \nonumber \\
&& +K*\log\left\{
\frac{(1+Y_{m-1}^{(a)})(1+Y_{m+1}^{(a)})}
{\prod_{d=1}^{s}(1+(Y_{m}^{(d)})^{-1})^{I_{ad}}}\right\}(v), 
\label{nonlinear}
\end{eqnarray}
where $Y^{(a)}_{0}(v)=0$, $a\in \{1,2,\dots,s\}$ and
 $m \in {\mathbb Z}_{\ge 1}$; 
$*$ is a convolution
\begin{eqnarray}
(f*h)(v)=\int_{-\infty}^{\infty}dw f(v-w)h(w),
\end{eqnarray}
and the kernel is 
\begin{eqnarray}
K(v)=\frac{1}{2 \cosh \pi v}. 
\end{eqnarray}
Substituting $u=-\frac{\beta {\mathcal J}}{N}$ and 
taking the Trotter limit $N \to \infty$, 
we obtain the TBA equation 
\begin{eqnarray}
\log Y_{m}^{(a)}(v)=
\frac{\pi {\mathcal J} \beta \delta_{ap}\delta_{mb}}{\cosh\pi v}
+K*\log\left\{
\frac{(1+Y_{m-1}^{(a)})(1+Y_{m+1}^{(a)})}
{\prod_{d=1}^{s}(1+(Y_{m}^{(d)})^{-1})^{I_{ad}}}\right\}(v), 
\label{TBA-2}
\end{eqnarray}
where $a\in \{1,2,\dots,s\}$, $m\in {\mathbb Z}_{\ge 1}$,
 $Y_{0}^{(a)}(v):=0$. This TBA equation (\ref{TBA-2}) is 
identical to the one from the string hypothesis. 
Taking note on the relations
\begin{eqnarray}
&& C_{ad}(v)=\sum_{l=1}^{\min(a,d)}G_{|a-d|+2l-1}(v) ,\nonumber \\
&& G_{a}(v)=\frac{4}{2s+1}
\frac{\cos\frac{(2s+1-2a)\pi}{4s+2} \cosh\frac{2\pi v}{2s+1}}
{\cos\frac{(2s+1-2a)\pi}{2s+1} + \cosh\frac{4\pi v}{2s+1}},\nonumber \\
&& \widehat{C}_{ad}(k)=\int_{-\infty}^{\infty}{\mathrm d}v
 C_{ad}(v)e^{-ikv}, \nonumber \\
&& \sum_{c=1}^{s}\widehat{C}_{ac}(k)\widehat{D}_{cd}(k)=\delta_{ad},
\label{relations} \\
&& \widehat{D}_{cd}(k)=2\delta_{cd}\cosh\frac{k}{2}-I_{cd},\nonumber
\end{eqnarray}
 one can also rewrite this TBA equation as
\begin{eqnarray}
\log Y_{m}^{(a)}(v)&=& 2\pi \beta {\mathcal J}
 \delta_{m1}G_{a}(v) \nonumber \\ 
&& +\sum_{b=1}^{s}C_{ab}*\log\left\{
\frac{(1+Y_{m-1}^{(b)})(1+Y_{m+1}^{(b)})}
{\prod_{d=1}^{s}(1+Y_{m}^{(d)})^{I_{bd}}}\right\}(v),
\label{TBA3}
\end{eqnarray}
where $Y^{(a)}_{0}(v)=0$, $a\in \{1,2,\dots,s\}$ and
 $m \in {\mathbb Z}_{\ge 1}$.
In contrast to (\ref{TBA-2}), (\ref{TBA3}) does not 
contain $1+(Y^{(a)}_{m}(v))^{-1}$ which is not relevant to 
evaluate the central charge  for the case ${\mathcal J}<0$. 
One can also derive the following relation 
from (\ref{T-system}) for $m=1$, (\ref{Y-fun}) and (\ref{relations}).
\begin{eqnarray}
\log T^{(1)}_{1}(v)&=&\log \phi_{-}(v+i)\phi_{+}(v-i) 
 +\sum_{a=1}^{s}G_{a}*\log(1+Y^{(a)}_{1}) 
 \nonumber \\
&& +N \int_{0}^{\infty}{\mathrm d}k 
 \frac{2e^{-\frac{k}{2}}\sinh(ku)\cos(kv)\cosh(\frac{2s-1}{4}k)}{
 k \cosh(\frac{2s+1}{4}k)}.
\end{eqnarray}
Taking the Trotter limit $N \to \infty$ with $u=-\frac{{\mathcal J}\beta}{N}$, 
we obtain 
the free energy density 
 ${\mathcal F}=-\frac{1}{\beta}\log T^{(1)}_{1}(0)$  
without infinite sum. 
\begin{eqnarray}
{\mathcal F}&=&{\mathcal J}\left\{
\frac{2}{2s+1}\left(
2\log 2 -\psi(\frac{1}{2s+1})+\psi(\frac{3+2s}{2+4s})
\right)
-1 
\right\} \nonumber \\
&&-k_BT
\sum_{a=1}^{s}
\int_{-\infty}^
{\infty}{\mathrm d}v G_{a}(v)\log(1+Y^{(a)}_{1}(v)), 
\label{free-finite}
\end{eqnarray}
where $\psi(z)$ is the digamma function
\begin{eqnarray}
\psi(z)&=&\frac{d}{dz}\log \Gamma (z).
\end{eqnarray}
The first term in the rhs of (\ref{free-finite}) for ${\mathcal J}=-1$ 
coincides with the grand state energy of
 the $osp(1|2s)$ model in \cite{Mar95-2}. 
 Using the result of this section, we can 
 show that the central charge of the corresponding system is $s$. 
%%%%%%%%%%%%%%%%%%%%%%%%%%%%%%%%%
%%%%%%%%%%%%%%%%%%
\section{Discussion}
In this paper, we have derived the TBA equation from the $osp(1|2s)$ 
version of the $T$-system. 
The $osp(r|2s)$ integrable spin chain is related to interesting physical 
problems, such as the loop model which is 
 related to statistical properties of 
polymers\cite{MNR98}, 
and the fractional quantum Hall effect \cite{HR88}, etc. 
So it is desirable to study the $osp(r|2s)$ integrable spin chain
 beyond the $osp(1|2s)$ case. 
For $r>0$ case,  
 we have only the $T$-system for tensor-like representations \cite{T99-2}. 
 To construct a complete set of the $T$-system which is relevant for 
 the QTM method, we have to treat spinorial representations.

%%%%%%
In closing this paper, we shall mention the $sl(r+1|s+1)$ 
version of the $T$-system \cite{T97,T98-1,T98-2} 
which is omitted in this paper. 
The $osp(1|2s)$ $T$-system is obtained as a reduction of 
a kind of Hirota-Miwa equation. 
This is also the case with $sl(r+1|s+1)$. 
For $m,a \in {\mathbb Z}_{\ge 1}$, $sl(r+1|s+1)$ $T$-system 
leads as follows. 
\begin{eqnarray*}
&& T_{m}^{(a)}(v-1) T_{m}^{(a)}(v+1)
=T_{m+1}^{(a)}(v)T_{m-1}^{(a)}(v)+
T_{m}^{(a-1)}(v)T_{m}^{(a+1)}(v) 
\\ 
&& \hspace{0pt} {\rm for} \quad 
1 \le a \le r \quad {\rm or} \quad 1 \le m \le s
\quad {\rm or} \quad (a,m)=(r+1,s+1),
\end{eqnarray*}
\begin{eqnarray*}
&& T_{m}^{(r+1)}(v-1) T_{m}^{(r+1)}(v+1)
=T_{m+1}^{(r+1)}(v)T_{m-1}^{(r+1)}(v) 
\quad {\rm for} \quad m \ge s+2, 
\end{eqnarray*}
\begin{eqnarray*}
&& T_{s+1}^{(a)}(v-1) T_{s+1}^{(a)}(v+1)
=T_{s+1}^{(a+1)}(v)T_{s+1}^{(a-1)}(v) 
\quad {\rm for} \quad a \ge r+2.
\end{eqnarray*}
where,
\begin{eqnarray*}
&& T^{(a)}_{s+1}(v)=\epsilon_{a} T^{(r+1)}_{a+s-r}(v) 
\quad {\rm for} \quad   
a \ge r+1, \nonumber \\ 
&& T^{(0)}_{m}(v)=T^{(a)}_{0}(v)=1.
\end{eqnarray*}
Here we omit the vacuum part which can be easily recovered 
so as to be compatible with the lhs (vacuum part) of the BAE. 
The phase factor $\epsilon_{a}$ depends on the definition 
of the transfer matrix. For example, if the transfer matrix 
is defined as a supertrace of a monodromy matrix, we have 
 $\epsilon_{a}=(-1)^{(s+1)(a+r+1)}$. 
Note that above functional equation 
reduces to the  $T$-system for $sl_{r+1}$ \cite{BR90} 
(see also \cite{KP92,KNS94-1}) 
if we set $s=-1$. 
%%%%%%
              
\end{document}